\documentclass[prd, twocolumn, nofootinbib, floatfix]{revtex4-2}
\pdfoutput=1  
\usepackage{amsmath}
\usepackage{graphicx}
\usepackage{dcolumn}
\usepackage{bm}
\usepackage{epsfig}
\usepackage{amssymb,latexsym,mathrsfs}
\usepackage{graphicx}
\usepackage{color}
\usepackage{hyperref}

\usepackage{float}

\hypersetup{
    colorlinks=true,
    linkcolor=red,
    citecolor=blue,
}

\newcommand{\be}{\begin{equation}}
\newcommand{\ee}{\end{equation}}
\newcommand{\bs}{\begin{split}} 
\newcommand{\bea}{\begin{eqnarray}}
\newcommand{\eea}{\end{eqnarray}}

\newcommand{\om}{\Omega_m}

\begin{document}

\title{What is the Standard Cosmological Model?} 

\author{Eric V.\ Linder$^{1,2}$} 
\affiliation{${}^1$Berkeley Center for Cosmological Physics \& Berkeley Lab, 
University of California, Berkeley, CA 94720, USA\\
${}^2$Energetic Cosmos Laboratory, Nazarbayev University, 
Nur-Sultan 010000, Kazakhstan
}

\begin{abstract} 
Reports of ``cosmology in crisis'' are in vogue, but as Mark Twain said, 
``the report of my death was an exaggeration''. We explore what we might 
actually mean by the standard cosmological model, how tensions -- or their 
apparent resolutions -- might arise from too narrow a view, and why looking at the 
big picture is so essential. This is based on the seminar ``All Cosmology, 
All the Time''. 
\end{abstract}

\date{\today} 

\maketitle

\section{Introduction}

The origin of the word ``crisis'' comes from ``decision'', implying a 
framework -- a standard model -- and paths forward that are being 
decided between. So to assess whether some crisis exists and how 
severe it might be, we must first 
explore what we mean by the standard 
cosmological model. 

We discuss various definitions in Section~\ref{sec:what}, then 
investigate a couple of oft claimed tensions and their (un)resolutions in 
Section~\ref{sec:tense}. The primacy of using all robust observations, 
at all redshifts -- All Cosmology, All The Time -- is motivated in 
Section~\ref{sec:all}, and the concluding discussion is in Section~\ref{sec:concl}.

\section{What is Standard Cosmology?} \label{sec:what} 

What is the standard cosmological model depends very much on where one 
draws the line on what is cosmology or the universe.

\subsection{Level 1: Global Properties} 

One could define Level 
1 as saying cosmology is the global properties of the universe: 
\begin{itemize} 
\item Connected -- a signal can get from there to here and from then 
to now. There is no discreteness (at least at this level). 
\item Metric -- we can figure out how far it is from there to here 
and from then to now. 
\end{itemize} 
This gives the foundation and few would dispute that they are part of the 
standard cosmological model; violation of either on the scale of the 
observable universe would be revolutionary (and pretty exciting on smaller 
scales as well!). 

A third property comes from a plethora of observations, and is also generally 
accepted as part of our standard cosmological model: 
\begin{itemize} 
\item Homogeneity and Isotropy -- on observable universe scales. 
\end{itemize} 
This then implies the Robertson-Walker metric and we are on familiar 
territory! 
In particular, the Robertson-Walker metric has two characteristics 
(independent of the theory of gravity): 
\begin{itemize} 
\item Evolution -- a scale factor $a(t)$. 
\item Spatial curvature -- a constant $k$ proportional to 
the Gaussian curvature of space. It is worth emphasizing that a 
Robertson-Walker spacetime can have spatial curvature but no spacetime 
curvature (though this does not describe our universe), and spacetime 
curvature but no spatial curvature (which fits observations pretty well). 
\end{itemize}

Unlike, say, discreteness, we don't have to go down to laboratory scales 
to test for a breakdown in homogeneity and isotropy, and hence the 
Robertson-Walker metric. However, one can show that these smaller scale 
deviations do not significantly affect cosmological scales. This is worth 
emphasizing: 
for the expansion and curvature, and for observations involving propagation 
of light rays along a line of sight, and for growth within three dimensional 
volumes, the effect from small scales generally is calculable and found to 
be small \cite{futamase89,lin90,jlw92,jlw93}.

\subsection{Level 2: History of the Universe}  

If one delves into the time structure, one could say that a Level 2 
answer to what is the standard cosmological model is to say cosmology 
is the history of the universe. From observations we could list five key 
stages: 
\begin{itemize} 
\item Early hot dense state -- popularly, if nebulously, known as the 
``Big Bang''. Here we consider it as a generic description and source 
of initial conditions, and whether it occurs at the Planck energy, 
$10^{15}$ GeV, or $10^3$ GeV is a (fascinating) detail, not a source of crisis. 
\item Matter/antimatter asymmetry -- I have absolutely nothing to say! 
While we know the basic ingredients to  deliver this (if not its magnitude) 
\cite{sakharov} one could well argue this is a greater crisis than any other 
raised, and yet it is rarely mentioned.  
\item Radiation dominated era -- primordial nucleosynthesis, degrees of 
freedom $g_\star$ (neutrino decoupling, electron/positron annihilation), 
CMB thermalization. Lots of good stuff! 
\item Matter dominated era -- CMB scattering, growth of structure (us!). 
\item Cosmic acceleration -- ``dark energy'', fate of the universe? 
\end{itemize} 

As an aside, 
one of the most fascinating aspects of this is that the argument could  
be made that the study of dark energy actually grew out of investigation of 
the radiation dominated era. The radiation era spans such a huge range 
of e-folds of expansion (nominally some 55, compared to the matter era's 
7, or dark energy's 0.5), and yet we know little of the details -- did 
it stay radiation dominated the entire time? We tend to simply assume this 
but we have accurate windows on 
only tiny slivers of this era, around primordial nucleosynthesis  
\cite{carroll,masso,dutta} and toward the end, just before matter domination 
and recombination 
\cite{1009.3500,1208.4845,hojjls}. In 1979, Robert Wagoner \cite{wag79} 
highlighted this by considering  what freedom there was of changing the 
equation of state of the dominant energy density. This program of exploring 
the equation of state at various epochs was picked 
up by two of his students and in the 1980s was applied to the 
late universe and became dark energy cosmology, 
developing both the cosmological model and observational probes, 
first for particular values of the equations of state 
\cite{turner83,turner87} and then for a 
general equation of state \cite{lin88a,lin88b}.

\subsection{Level 3: Stuff {\it In\/} the Universe} 

A more specific view is that cosmology is the stuff {\it in\/} the  
universe. After all, this is what is actually observed. This would 
include: 
\begin{itemize} 
\item Cosmic microwave background radiation (CMB) -- CMB structure (anisotropies, 
polarization, spectral distortions) is a rich probe of both history 
(Level 2, including initial conditions such as adiabatic perturbations) 
and the other contents (Level 3, e.g.\ matter stuff through scattering, 
gravitational potentials). And of course it provides strong evidence 
through isotropy (and homogeneity through the CMB felt by distance 
objects) for Level 1 cosmology. 
\item Large scale structure -- The continuous fields of matter: the 
density field, velocity field, acceleration (gravity) field -- generally as 
probed by individual sources. While  
these fields are related, the relations do test the framework and each has 
particular incisive elements so they can be considered distinct; plus, for cosmological 
purposes they are at very different stages of observational development. 
\item Other -- As probes of the standard cosmological model, observations 
of other stuff such as neutrinos, gravitational waves, exotica (e.g.\ 
topological defects) have not yet reached the same stage of having a major impact, 
though this would be an exciting development. 
\end{itemize}

\subsection{Turtles All the Way Down} 

Some researchers extend the standard model of cosmology further and 
further down in detail, to ``the stuff in the stuff in the universe'', 
e.g.\ aspects of galaxies, galaxy clusters, generation of various 
particles and fields (e.g.\ neutrinos, gravitational waves, etc.). 
Others will stretch to ``the properties of the stuff in the stuff in 
the universe'', e.g.\ cuspy cores of galaxies, tidal streams, Cepheid 
pulsations, etc. 

Where one draws the line between the standard cosmological model and 
all else -- call it astrophysics for simplicity -- is a personal choice. 
But it is rare that rain next Tuesday in some place one had not predicted it 
throws the standard meteorological model into crisis. It is not impossible, 
but it is useful to remember that some particular tension has  
to work its way from Level $N$ to the fundamental foundations.

\subsection{Cosmologing Is Hard} \label{sec:cosmologing} 

But\dots cosmologing is hard. The properties of the stuff in the stuff 
affect how and what we learn about the more fundamental stuff. We are 
then faced with the puzzle of whether we have fully understood the stuff 
in the stuff (or mismeasured its properties) or whether indeed it 
propagates cleanly to impacting the standard cosmological model. Let us 
take two examples. 

{\it Example  1.\/} Suppose one measured the CMB temperature at scale 
factor $a$ (redshift $z=a^{-1}-1$) and found that 
\be 
T_{\rm CMB}(z)\ne T_{\rm CMB}(0)\times(1+z)\,? 
\ee 
What should one conclude: that the universe is not adiabatically 
expanding, or that there is some systematic error (e.g.\ molecular 
collisional excitations)? How much effort, and what proportion of 
the literature, should be dedicated to investigating systematics before 
declaring a crisis in the standard cosmological model? 

{\it Example 2.\/} Suppose one measured the distance to an object (or set 
of objects) at redshift $z$ in terms of both its luminosity distance 
and angular diameter distance and found that the reciprocity relation is 
broken, 
\be 
d_L(z)\ne d_a(z)\times (1+z)^2\,? 
\ee 
If one wishes to conclude that this places the standard cosmological model 
in crisis (rather than some systematic error in the data), 
one must give up some foundational element, since this relation 
arises from a) metricity, b) geodesic completeness, c) photons propagate 
on null geodesics, and d) adiabatic expansion. (Note conservation of 
photon phase space density is a big part of this, but not all.) 

What level of systematics investigation should one (and the research 
community) carry out before 
declaring an upending of the standard cosmological model? To what extent 
should the proportion of systematics vs new physics papers depend on the 
degree of new physics required? As the saying goes, if you tell me you 
saw out the window a deer on the lawn, I might be willing to consider 
how the deer got there, its effect on the shrubbery, etc., but if you tell me 
you saw out the window a unicorn on the lawn, I might want more evidence 
before devoting time to the puzzle. 

We have in place solutions for how to handle conflicting expectations, 
observations, and theories: 
\begin{itemize} 
\item Rigorous data 
\item Multiple, disparate probes 
\item Crosschecks 
\item Consistency at all cosmic times 
\item Check the cosmic expansion history, cosmic growth history, and 
light propagation (and soon gravitational wave propagation) 
\end{itemize} 
We explore how these might be applied to an example tension in the 
next section.

\section{Past Tense, Present Tense?} \label{sec:tense} 

A well known tension lies in current Hubble constant $H_0$ values deduced 
from certain probes, taking all the data at face value. In addition to 
the main puzzle, there are some beyond the surface: 
\begin{itemize} 
\item Local measurements differ by some $\sim2\sigma$ depending on 
method, i.e.\ Cepheids vs tip of the red giant branch 
\cite{riess20,freedman20}. 
\item The tension is emphatically not ``early vs late'' cosmology since 
baryon acoustic oscillations (BAO) distances (together with primordial element 
abundances \cite{desbbn1,bbn2,bbn3} or marginalizing over or sidestepping 
the sound horizon at the baryon drag epoch \cite{drag,sherwin}), i.e.\ 
without use of the primordial CMB, gives the 
same answer as from the CMB. 
\item Strong lensing time delays show a sharp transition between low 
and high $H_0$ values around $z\sim0.4$ \cite{millon,liao}, albeit 
with a small sample. 
\end{itemize} 

\begin{figure}[!htb]
\centering 
\includegraphics[width=0.76\columnwidth]{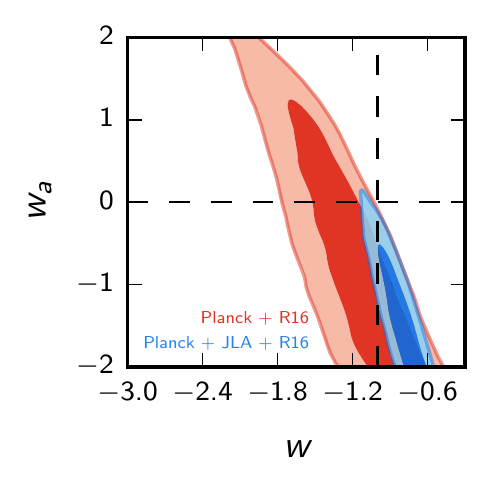}
\includegraphics[width=0.76\columnwidth]{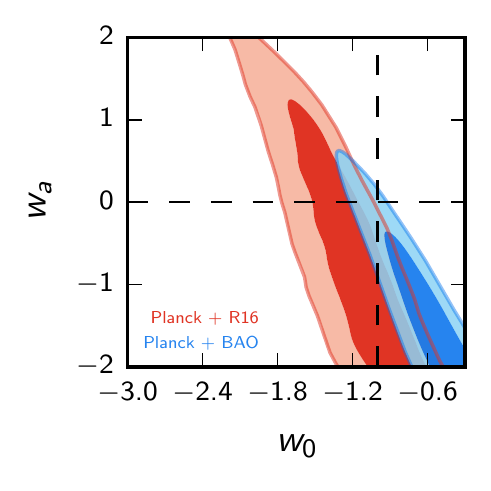} 
\includegraphics[width=0.76\columnwidth]{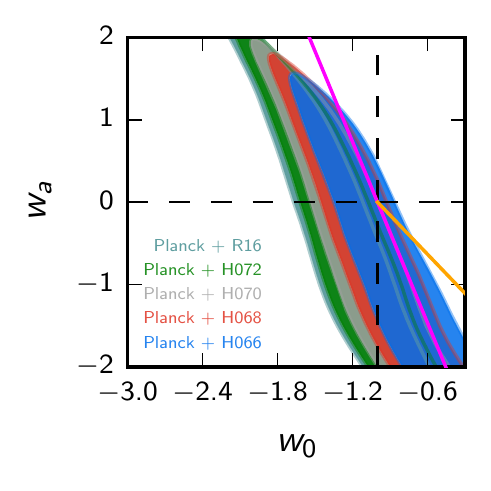} 
\caption{68.3\% and 95.4\% constraints on the $w_0$--$w_a$ plane in an 
$11$ parameter extended space, using Planck CMB data plus the R16 $H_0$ 
($\sim74$) prior. The top panel includes as well JLA supernova data, the middle panel 
includes baryon acoustic oscillations data: both strongly prefer $w_0\ge-1$. 
The bottom panel shows that $w_0\ge-1$ is most cleanly achieved by shifting 
the $H_0$ prior to $H_0\le70$. (Adapted from \cite{dival17}.) 
} 
\label{fig:h070} 
\end{figure}

While CMB data alone constrains $H_0$ tightly only within a $\Lambda$CDM 
cosmological model, while allowing a considerable range of $H_0$ when 
the dark energy equation of state $w$ differs from $-1$, it is 
extraordinarily difficult from a combination of cosmic probes such as 
CMB+BAO or CMB+SN (supernovae distances) to obtain $H_0>70$ 
\cite{dival17} (we always write $H_0$ in units of km/s/Mpc). Basically, 
$H_0>70$ requires a phantom dark energy ($w<-1$), which is disfavored 
by the above combinations of probes; see Figure~\ref{fig:h070}.

There are two basic loopholes one might try by changing the cosmic expansion 
history -- relax the tension in the present or the past. 
\begin{itemize} 
\item Present tense (late time transition): arrange a very sharp phantom 
excursion very close to the present so that higher redshift distances are 
not too strongly affected. 
\item Past tense (early time transition): arrange a lower sound horizon 
scale $r_{\rm drag}$ with the Hubble parameter $H$ going up. Again, one 
must make it a sharp transition -- a spurt of extra early energy density 
to raise $H$, then removing the early dark energy to preserve the agreement 
with CMB data. 
\end{itemize} 

\begin{figure}[htb!]
\centering 
\includegraphics[width=\columnwidth]{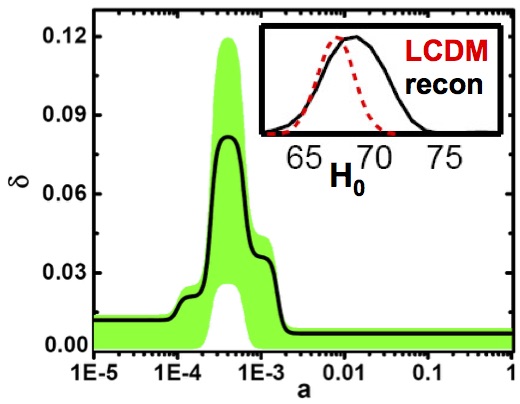}
\caption{Reconstruction of the expansion history deviations 
$\delta(a)=\delta H^2/H^2_{\Lambda{\rm CDM}}$ from $\Lambda$CDM is 
shown, with the mean value (solid line) and 68\% uncertainty band 
(shaded area). Note the data prefers early dark energy $\delta\ne0$. 
The inset demonstrates that the model independent reconstruction then prefers a 
higher $H_0$ than the $\Lambda$CDM standard analysis. 
(Adapted from \cite{hojjls}.) 
} 
\label{fig:hls} 
\end{figure}

We begin with the early time transition. The covariance between $r_{\rm drag}$ 
and $H_0$ has been known for a long time 
\cite{bondef98,eiswh04,doran,robbers,hojjls}. Using 
CMB data, in 2013 Ref.~\cite{hojjls} actually found evidence for an early time 
transition and its effect on $H_0$! -- see Figure~\ref{fig:hls}. 
This has been resuscitated in many 
many articles in the last couple years. However, early time transitions 
do not really work in removing the Hubble constant tension (see, e.g., 
\cite{knox,2003.07355,2010.04158,caldwell}). On the one hand they cannot viably raise 
$H_0$ as far as 
the high values favored by Cepheids, and on the other hand they generically 
violate other aspects of the cosmological model as we discuss in 
Section~\ref{sec:all}.

The late time transition has the advantage that there is much less effect 
on the CMB, and so more freedom for change. If one raises $H(z)$, distances 
will change. To preserve distances (e.g.\ the distance to the CMB last 
scattering surface, as well as BAO and SN distances), with a higher $H_0$ 
one needs a smaller $H(z>0)$. This means less energy density. This could 
be from a smaller matter density $\om$ (the matter density today as a 
fraction of the critical density) but this is insufficient and one requires 
a smaller dark energy density at some $z>0$. To give enough dark energy 
density today (especially if the matter density is low and the total density 
is the critical density) requires dark energy density to appear quite 
suddenly at low redshifts. This is again the phantom regime $w<-1$. 
And again, such late time phase transitions have been known for a long 
time (see, e.g., vacuum metamorphosis \cite{parkerr00,parkerv04,caldwell06}). 

Since such late time transitions all have basically the same physical effect, 
regardless of specific origin, let's examine whether vacuum metamorphosis 
removes the $H_0$ tension. Yes! but\dots. As the opening lines of the 
abstract of \cite{dival20} say, ``We do obtain $H_0\approx 74$ km/s/Mpc 
from CMB+BAO+SN data in our model, but that is not the point.'' This -- 
and essentially any late time transition -- model fails because there is 
more to the standard cosmological model than $H_0$; it does not satisfy 
other probes. We discuss the details in Section~\ref{sec:all}. 
For other possible issues with late time transitions see 
\cite{knox,2002.11707,2101.08641,2103.08723}.

\section{All Cosmology, All The Time} \label{sec:all} 

So far we have considered only the expansion history. However, one must 
take into account all cosmological probes, e.g.\ how deviations from a 
standard cosmological model affect the growth of large scale structure. 

In the vacuum metamorphosis case of the previous section, the combination 
of probes CMB+BAO+SN produced $H_0\approx74$. For a good fit to the CMB, 
preserving $\om h^2$ means a low $\om\approx0.27$. That can be ok. 
However, it also gives a high amplitude for mass fluctuations, 
$\sigma_8\approx0.88$, which is quite high. This is due to the reduced 
dark energy density needed to get the distances right, quite generally 
implying greater 
matter domination and growth at higher redshifts. We can start to see 
that we do indeed need ``all cosmology, all the time'' -- use of all 
probes, over the full cosmic history. 

\begin{figure}[htb!]
\centering 
\includegraphics[width=\columnwidth]{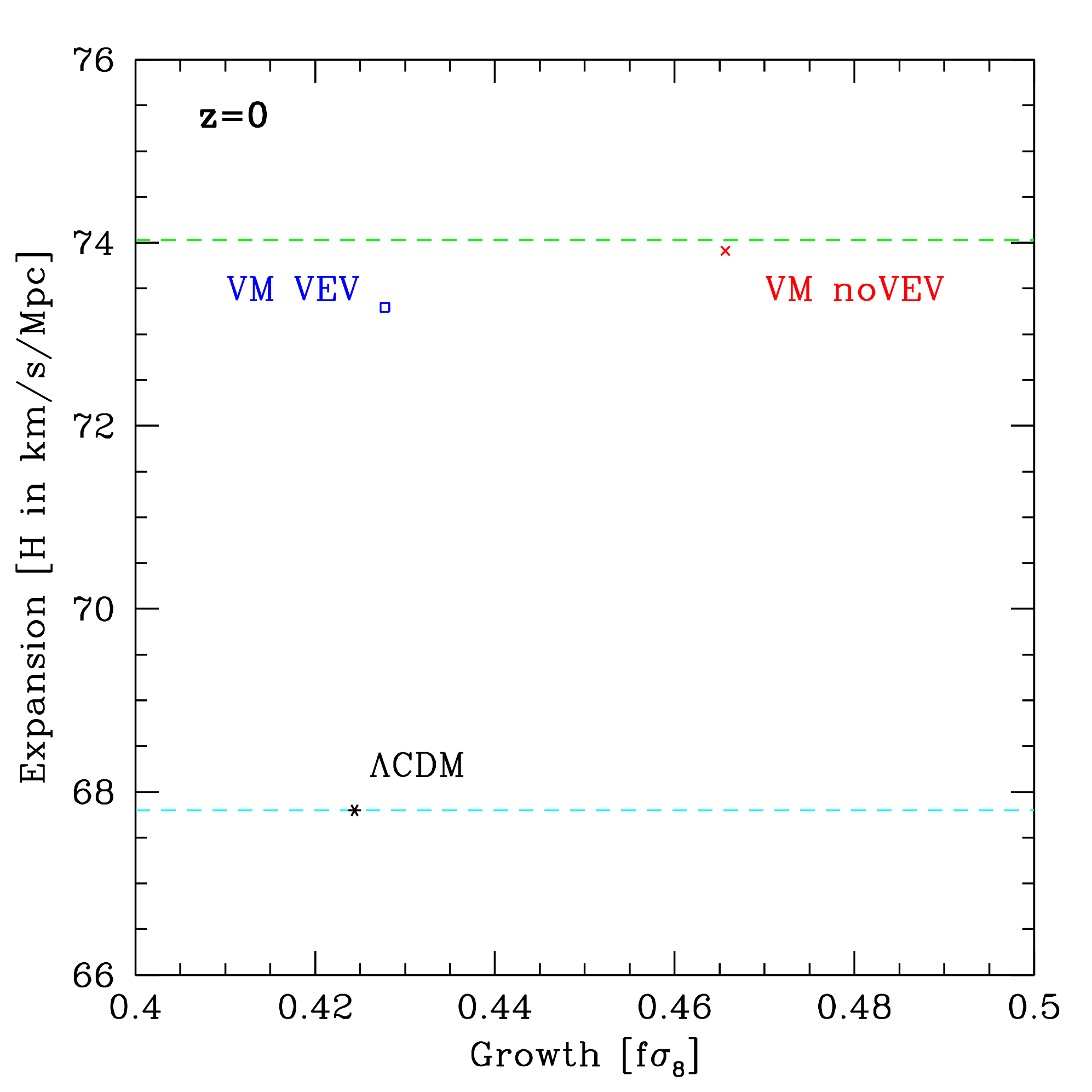} 
\caption{Expansion and growth histories are here plotted simultaneously. 
At $z = 0$, hence these give $H_0$ and $f\sigma_8(0)$ ($\approx 0.52\,S_8$ 
for most viable cosmologies in general relativity; see \cite{dival20}), shown 
by the points. 
Excellent agreement with the Cepheid value of $H_0$ 
(green dashed line), as opposed to the $\Lambda$CDM value (cyan 
dashed line) is obtained by the flat vacuum metamorphosis (VM) late 
transition models with parameters fit to CMB+BAO+SN, and the VM 
VEV model agrees well on $S_8$ as well. 
(Adapted from \cite{dival20}.) 
} 
\label{fig:expgro0} 
\end{figure}

One might brush high $\sigma_8$ under the rug and say that with the lower 
$\om$, one has $S_8\equiv\sigma_8(\om/0.3)^{0.5}\approx0.83$, which might 
be workable for some probes, i.e.\ roughly as good as $\Lambda$CDM. So 
we could say that vacuum metamorphosis gives $H_0\approx 74$ while not 
making any $S_8$ tension worse, as apparently seen in 
Figure~\ref{fig:expgro0}. 

\begin{figure}[htb!]
\centering 
\includegraphics[width=\columnwidth]{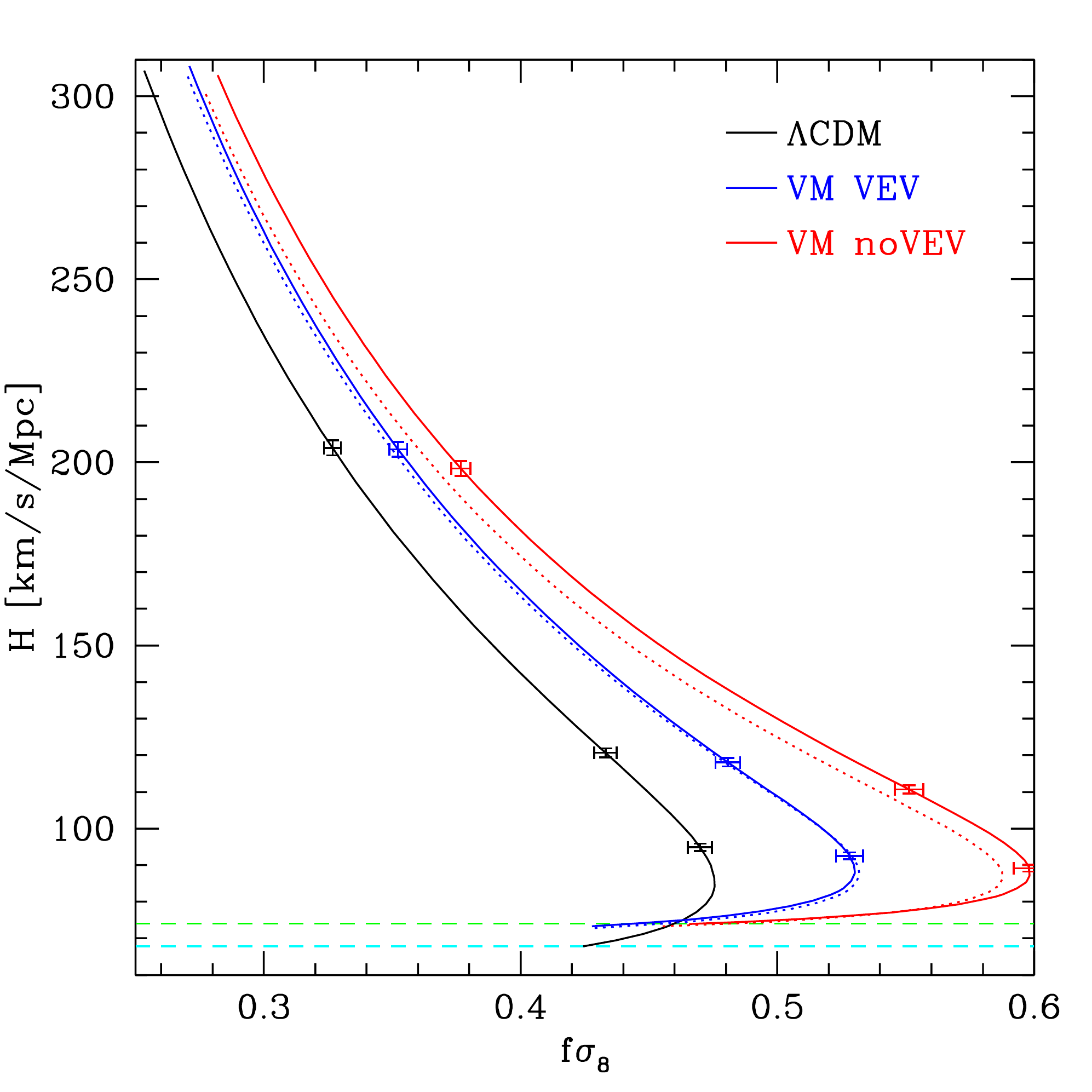} 
\caption{A very different situation occurs if one looks 
beyond $H_0$ at the full conjoint evolutionary track of the various models, 
not just the $z = 0$ (lower) endpoints as shown in Figure~\ref{fig:expgro0} 
(which covers the small range between the dashed lines). 
Over the histories the VM models diverge considerably from $\Lambda$CDM. 
Curves extend from $z = 0$ at the bottom to $z = 3$ at the top, and the points 
with error bars show the trajectory location at $z = 0.6$, 1, 2, where the error 
bars mimic 1\% constraints on each axis quantity to give a sense of separation 
between the curves. Solid curves are for flat space, dotted curves include 
$\Omega_k$. 
(Adapted from \cite{dival20}.) 
} 
\label{fig:expgroz} 
\end{figure}

However, $S_8$ and $H_0$ are focusing on a single time (the present) in 
cosmic history. This is a Bad Idea. In the words of Lewis Carroll, ``the 
rule is, jam tomorrow and jam yesterday -- but never jam today''. 
Figure~\ref{fig:expgroz} shows that the apparent removal of the $H_0$ 
tension is moot, since Figure~\ref{fig:expgro0} is only a tiny part of 
cosmic history, and when one does all cosmology, all the time -- taking 
into account both the cosmic expansion and cosmic growth over 
the span of cosmic history as in Figure~\ref{fig:expgroz} -- then 
generically late time transitions do not work in giving a viable 
cosmological model.

This is general because the late time transition, 
anchored by the present, requires a lower dark energy density at higher 
redshifts and hence greater growth. If you squeeze the model in one place, 
it will bulge out elsewhere and fail to fit the array of probes. 
A similar general ``no-go'' reasoning holds for early time transitions. The higher 
expansion rate damps the CMB perturbations, so to preserve the fit to CMB 
data one requires a higher primordial curvature perturbation amplitude, 
and this strengthens the later growth of the matter perturbations, leading 
to high $\sigma_8$. (See also \cite{2103.04045}.)

\section{Discussion} \label{sec:concl} 

Very generally, neither late time nor early time attempts to remove 
the $H_0$ tension survive all cosmology, all the time. 
One needs to take into account all the probes, at all redshifts. 
It is not just $H_0$, it is $H(z)$. It is not just $\om$, it is $\om(z)$, 
and growth of structure, light propagation, etc. 

The standard cosmological model, whether one views it at Level 1, 2, or 
whatever, is strong: I come to praise it, not to bury it. 

But, {\it arguendo\/}, suppose we ignored the questions of 
Section~\ref{sec:cosmologing} and believed all data blissfully free of 
systematics, so that the order (two orders?) of magnitude difference 
in number of papers squeezing the standard cosmological model vs 
investigating the data is right and proper. What could we do to remove 
the tension? From the preceding section it is clear: we must break the 
connection between the cosmic expansion history and growth history. 
We can do this by a breakdown in general relativity, making gravity 
weaker so that $\sigma_8$ becomes lower; we can do this by introducing 
new particle interactions, again to suppress growth. Such changes will in 
turn alter other cosmological probes, which must be considered, 
and we must be wary of a spiral 
of epicycles. (See also \cite{2103.01183}.)  

Basically we would need to break the standard expansion history to address 
the $H_0$ tension, and need to break the standard growth history to 
keep the consequences of the first break consistent with other probes. 
An interesting possibility is to probe the connection of the growth of structure to the 
cosmic expansion in a wholly new way. Gravitational wave standard sirens 
offer this possibility, in part. A direct connection between siren 
distances and matter growth was developed by \cite{noslip}. 

The propagation of gravitational waves is an excellent probe of ``spacetime 
friction'' -- a combination of the Hubble friction from expansion and any 
time variation of the gravitational strength; the growth of structure probes 
both these quantities as well, in a different way. By comparing these two probes, 
possibly plus light 
propagation which depends only on the expansion, we have the potential to 
get a clear view of whether the connections between them in the standard 
cosmological model hold -- and if they do not, is there consistency between 
the deviations in one probe and in another. 

A bit more technically: while distances found through light propagation 
$d_{EM}(z)=d_L(z)$ depend on the expansion $H(z)$, those determined through 
gravitational wave propagation depend on both $H(z)$ and 
$\alpha_M(z)\equiv d\ln M^2_{Pl}(a)/d\ln a$, the running of the Planck mass or 
inverse gravitational strength. Thus, a measurement $d_{GW}(z)\ne d_{EM}(z)$ 
would signal a deviation from general relativity (or systematics). 
Meanwhile, growth of cosmic structure depends on $H(z)$ and the strength 
of gravity $G_{\rm eff}(z)$. In theories where the gravitational strength 
is determined purely by the Planck mass, $G_{\rm eff}(z)=1/M_{Pl}^2(z)$, 
then the circle is complete and there is a tight relation between the 
three probes. (Some theories of gravity have a further factor, the 
braiding that mixes the tensor and scalar parts, loosening the relation; 
$G_{\rm eff}(z)$ can also affect light propagation distances, and hence $H_0$ 
from strong lensing, giving a redshift dependence to the derived $H_0$ \cite{liao}.) 
Thus we can crosscheck against systematics by seeing if the relation 
holds: a deviation in one probe {\it predicts\/} a specific redshift 
dependence for a deviation in another probe. 

\begin{figure}[!tbh]
\centering 
\includegraphics[width=\columnwidth]{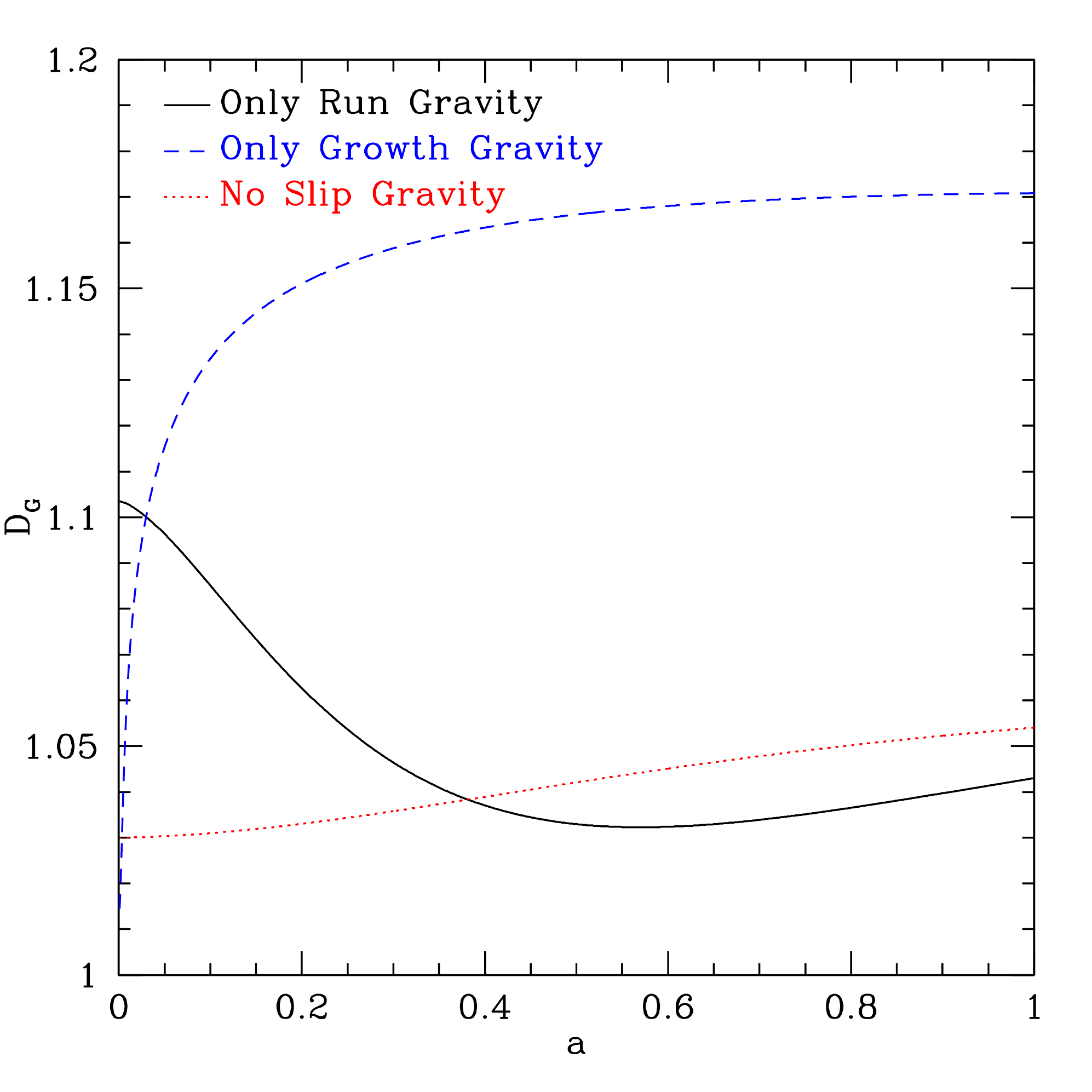} 
\caption{The new $D_G$ statistic, using the complementarity 
of the gravitational wave luminosity distance $d_{L,GW}$ and the cosmic 
matter growth rate $f\sigma_8$, can clearly distinguish different classes 
of gravity. Each class has a distinct shape in its redshift dependence 
$D_G(a)$, though the curve amplitudes will scale with $G_{\rm eff}(z=0)$. 
General relativity has constant $D_G = 1$. 
(Adapted from \cite{lmg}.) 
} 
\label{fig:dg} 
\end{figure}

The consistency check can be quantified with a new statistic combining 
the probe measurements \cite{lmg}, 
\be 
D_G(a)\equiv \frac{\left[d_{L,GW}^{\rm MG}/d_L^{\rm GR}\right](a)}{\left[f\sigma_8^{\rm MG}/f\sigma_8^{\rm GR}\right](a)}\,. 
\ee 
For general relativity, this equals one for all redshifts. For a given 
modified gravity (MG) model, it has a {\it specific\/} redshift 
dependence predicted. Several examples are shown in Figure~\ref{fig:dg}. 

In summary, the standard cosmological model has extremely deep foundations 
and multiple layers, and apparent surface blemishes may have very little 
to do with the fundamental basis. Especially if tensions cannot be solved 
by one ``tooth fairy'', such as an early or late time transition, when 
confronted with application of the principle of using all cosmology, all 
the time, but we are instead led to a series of epicycles, we might recall 
the unicorn vs the deer and pause.

New probes and data covering more cosmic history will be essential in 
exploring the standard cosmological model, at whatever level we define 
it, and whether it will stand firm, need some patchwork, or be overturned. 
Each of the levels has a wonderful array of areas for students to 
research, and make significant and lasting contributions to what {\it their\/} 
students will learn as the standard cosmological model.

\acknowledgments 

I thank the Asia-Pacific Center for Theoretical Physics for inviting 
me to give this inaugural lecture in June 2020 for the series 
``Dark Energy in a Dark Age''. 
This work is supported in part by the Energetic Cosmos Laboratory and by the 
U.S.\ Department of Energy, Office of Science, Office of High Energy 
Physics, under contract no.~DE-AC02-05CH11231.


\end{document}